\documentclass[a4paper,fleqn,usenatbib,useAMS]{mnras}

\usepackage{graphicx}	
\usepackage{amsmath}	
\usepackage{amssymb}	
\usepackage{multicol}        
\usepackage{bm}		
\usepackage{pdflscape}	
\usepackage{natbib}
\usepackage{comment}
\usepackage{newtxtext,newtxmath}

\title[The very short period young M6 dwarf binary system]{2MASS J10274572+0629104: The very short period young M6 Dwarf Binary System identified in K2 data}

\author[R. R. Paudel]{R. R. Paudel$^{1}$\thanks{Contact e-mail: \href{mailto:rpaudel@udel.edu}{rpaudel@udel.edu}}, J. E. Gizis$^{1}$, A. J. Burgasser$^{2}$, C. Hsu$^{2}$ 
\\
$^{1}$Department of Physics and Astronomy, University of Delaware, Newark, DE, 19716, USA\\
$^{2}$Center for Astrophysics and Space Science, University of California San Diego, La Jolla, CA 92093, USA}

\pubyear{2018}

\begin{document}
\label{firstpage}
\pagerange{\pageref{firstpage}--\pageref{lastpage}}
\maketitle

\begin{abstract}
We report the identification of a very low mass new binary system 2MASS J10274572+0629104, based on $Kepler$  $K2$ photometry and {\it Gaia} DR2 astrometry. The $K2$ light curve is consistent with a beat pattern of two periodic signals, and using Lomb-Scargle periodogram, we find two rotation periods of 0.2114$\pm$0.0002 day and 0.2199$\pm$0.0003 day. We conclude that these rotation periods arise from two stars with similar spectral types of M6, and have near equal luminosity. It is the first ultracool binary system to be identified based on beat patterns in the light curve. Near-infrared spectroscopy yields RV = -9.8$\pm$0.6 km s$^{-1}$, $v$ sin $i$ = 21.5$\pm$1.1 km s$^{-1}$, $T_{eff}$ = 3110$\pm$40 K, and log $g$ = 5.2${\pm}0.2$.  The motions are consistent with a young age, as are the rotation periods, but the source does not appear to be part of any known moving group. Furthermore, we detected three strong white light flares in the $K2$ light curve, with estimated total bolometric (UV/optical/IR) energies to be 2.8 $\times$ 10$^{33}$, 5.2 $\times$ 10$^{33}$ and 3.8 $\times$ 10$^{33}$ erg respectively.
\end{abstract}




\newpage

\section{Introduction}
M dwarfs, commonly known as red dwarfs, are the most populous stellar objects in the solar neighborhood and our Galaxy, comprising $\sim$75\% of all main sequence stars \citep{2017AJ....154..124C}. In addition, $\sim$25\% of these red dwarfs are found to be in binary or higher order multiple systems \citep{2013ARA&A..51..269D}, and ultracool dwarfs (with spectral type $\geq$M6) have a binary fraction of 20$\pm$4\% \citep{2007ApJ...668..492A}. The discovery of TRAPPIST-1 planetary system \citep{2016Natur.533..221G,2017Natur.542..456G,2017NatAs...1E.129L} demonstrates that M dwarfs have a high potential of hosting planets, making the study of various fundamental parameters such as rotational period, multiplicity, age, etc, important for characterizing these planets. The rotational period, in particular, measures the angular momentum evolution of M dwarfs, and can be used for gyrochronology - when combined with other properties like activity and color. Rotation also plays a role in powering the magnetic dynamo of fully convective stars with masses $M$ $<$ 0.35$M_{\odot}$ (\citealt{2016ApJ...821...93N} and references therein).    \\ \\
There are different methods of identifying binary systems, including direct imaging, radial velocity  (RV) variability, spectral blend inversion, astrometric variability and overluminosity on color magnitude diagrams (\citealt{2014ApJ...794..143B} and references therein). These methods are biased toward different types of binary systems based on separations, mass ratios and/or luminosity ratios. The high precision photometry of the $Kepler$ and $K2$ missions \citep{2010ApJ...713L..79K,2014PASP..126..398H} have provided another binary detection method: beat patterns in the combined light curves of variable stars with different rotation periods. Such patterns were seen in \textit{Kepler} light curves of rapidly rotating M dwarfs studied by \cite{2014ApJ...788..114R}. Here we report the discovery of a very low-mass binary by this method: 2MASS J10274572+0629104 (hereafter 2M1027+0629)  which most likely consists of two mid-M dwarfs with very similar rotation periods. \\ \\
2M1027+0629 (aka SDSS J102745.73+062910.1) was previously reported as a single M6 red dwarf by \cite{2011AJ....141...97W}. It has a high quality SDSS spectrum which shows H$\alpha$ emission with an equivalent width of 9.6$\pm$1.4 \AA. In Section \ref{sec: observations}, we describe $K2$ photometry, {\it Gaia} DR2 astrometry and photometry, NIRSPEC spectroscopy of 2M1027+0629, and white light flares observed on it. In Section \ref{sec: discussion} we discuss the results of observations obtained by $K2$, {\it Gaia} and NIRSPEC.
\section{observations} \label{sec: observations}
\subsection{\textit{K2} photometry}
\begin{figure*} 
    \includegraphics[scale=0.45,angle=0]{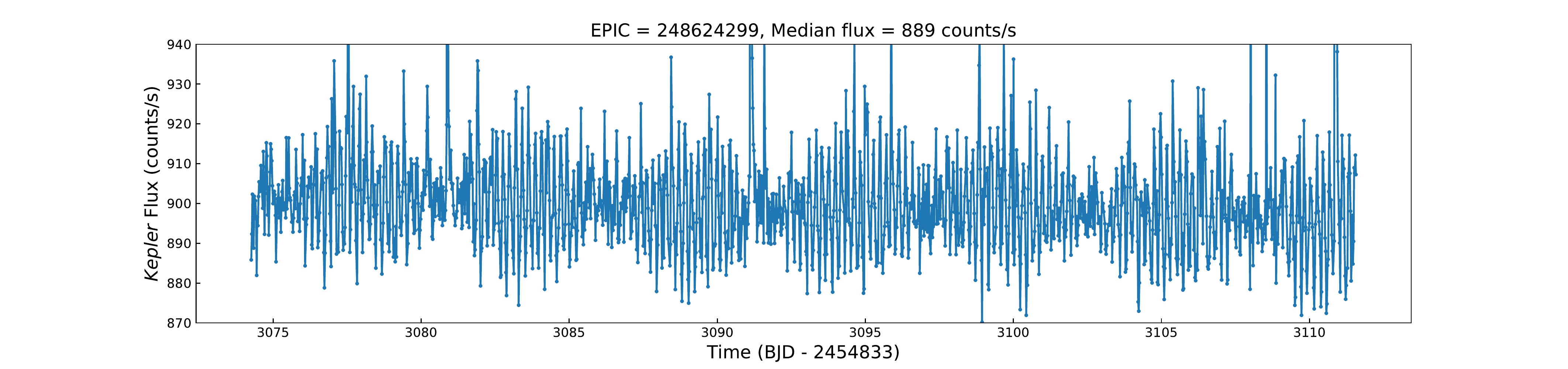} 
    \caption{$K2$ light curve of 2M1027+0629, in Campaign 14. Only a portion of the light curve is shown to focus on the beat patterns which are produced as a result of two closely separated periods 0.2114 day and 0.2199 day.}
    \label{fig:light_curve}
\end{figure*}

\begin{figure} 
    \includegraphics[scale=0.60,angle=0]{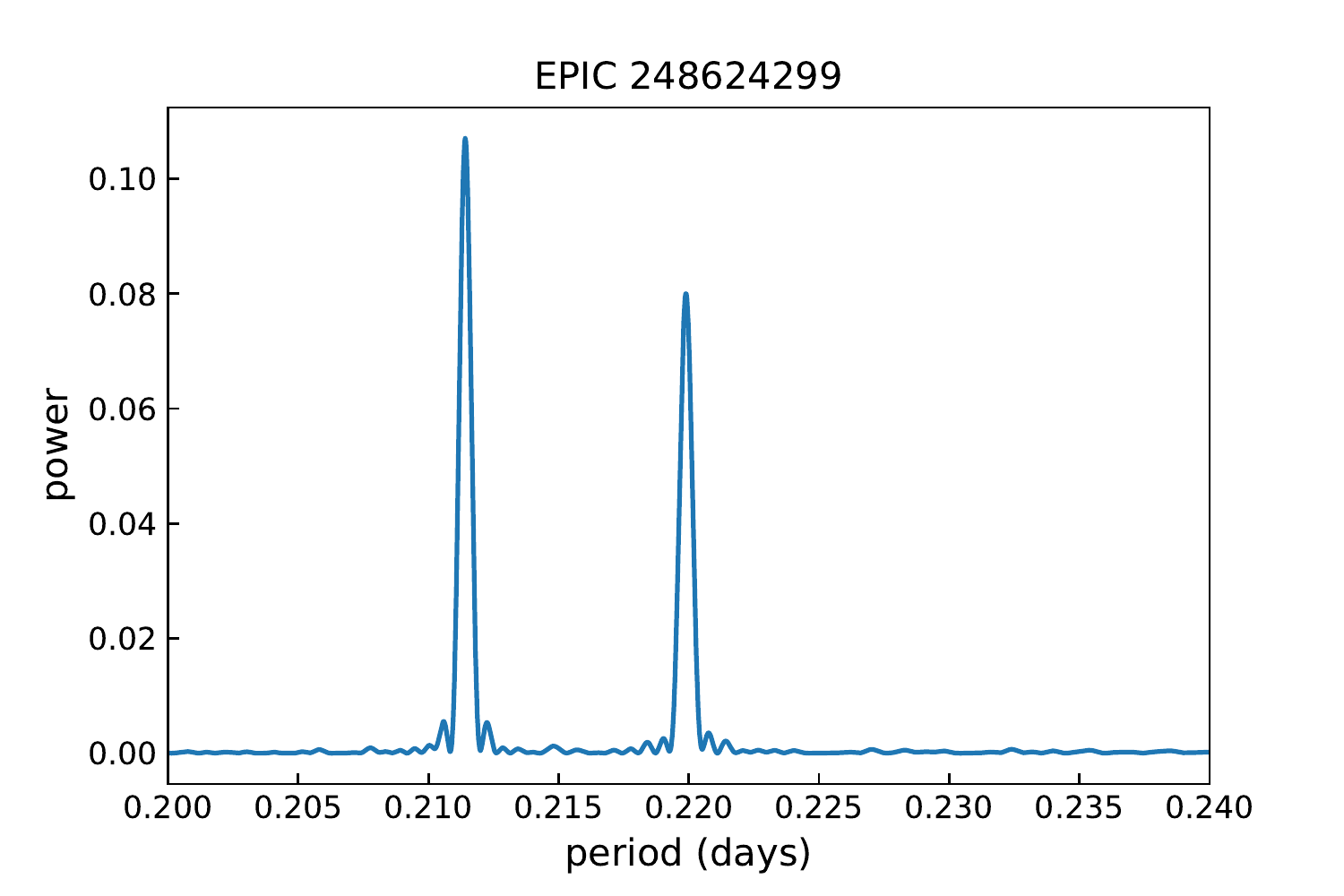} 
    \caption{Lomb-Scargle periodogram of 2MJ1027+0629. The two peaks correspond to periods 0.2114 days and 0.2199 days.}
    \label{fig:lomb_scargle_periodogram}
\end{figure}

2M1027+0629 was monitored continuously by $Kepler$ $K2$ mission \citep{2014PASP..126..398H} in long cadence ($\sim$30 minutes, \citealt{2010ApJ...713L.120J}) mode in Campaign 14 (31 May, 2017 - 19 August, 2017). Its Ecliptic Plane Input Catalog (EPIC) ID number is 248624299. The total observation time is 79.64 days, and the total number of good quality (Q = 0) data points is 3560. Each data point represents the average flux measured during a 29.4 minute interval. The median count rate through 2-pixel radius aperture is 889 counts s$^{-1}$. A part of $K2$ light curve is shown in Figure \ref{fig:light_curve} in which we can see the beat patterns resulting from two closely separated periods. The Lomb-Scargle (LSP) periodogram of the $K2$ light curve is shown in Figure \ref{fig:lomb_scargle_periodogram} with the power in $Y$-axis expressed in some arbitrary units. The LSP is constructed by taking a sample of 10,000 periods in the range (0.1-0.32)day, after getting an initial estimation that the periods lie in this range. The two closely spaced peaks in this LSP correspond to 0.2114$\pm$0.0002 days and 0.2199$\pm$0.0003 days, which are most likely due to spot modulations of two rapidly rotating stars in a binary system.  The uncertainties in the periods are based on half width at half maximum (HWHM) of the periodogram peaks as suggested by \cite{2013AJ....145..148M}. 
\begin{table}
 	\caption{Properties of 2M1027+0629}
 	\label{table:properties}
     \centering
     \scalebox{0.7}{
     \begin{tabular}{cccc}
     \hline
     \hline
       & Value & Units & Ref. \\
       \hline
        PHOTOMETRIC PROPERTIES \\
       \hline
       Sp. Type & M6  &  & 1 \\
      \textit{J} & 14.11$\pm$0.03 & mag & 1 \\
      \textit{H} & 13.51$\pm$0.04 & mag & 1 \\
      \textit{K$_{s}$} & 13.22$\pm$0.04 & mag & 1 \\
      \textit{ i }& 17.034$\pm$0.004 & mag & 2 \\
      \textit{G} & 17.782$\pm$0.003 & mag & 3 \\
      \textit{K$_{p}$} & 18.17 & mag & 4 \\
      H$\alpha$ EW & 9.6$\pm$1.4 & \AA & 1 \\
      \hline
      \hline
      PHYSICAL PARAMETERS\\
       \hline
       \hline
        distance (phot) & 67.6 & pc & 1 \\
        $\alpha$ & 156.940573$^{a}$ ($\pm$0.2 mas) & deg & 3 \\
        $\delta$ & 6.486120$^{a}$ ($\pm$0.3 mas) & deg & 3 \\
        parallax & 10.1 $\pm$ 0.3 & mas & 3\\
        $\mu_{\alpha}$ & -1.2 $\pm$ 0.5 & mas yr$^{-1}$ & 3 \\
         $\mu_{\delta}$ & -15.1 $\pm$0.5 & mas yr$^{-1}$ & 3 \\
         \hline
          \hline
          SPECTRAL PROPERTIES \\
          \hline
          \hline
           RV   & -9.8$\pm$0.6 & km s$^{-1}$ &  \\
            $v$ sin$i$ & 21.5$\pm$1.1 & km s$^{-1}$ & \\
            T$_{eff}$ & 3110$\pm$40 & K & \\
            log {\it g} & 5.2$\pm$0.2 & cgs & \\
             \hline
             \hline
             CALCULATED KINEMATICS\\
             \hline
             \hline
             $X$ & -34.1$\pm$1.1 & pc \\
             $Y$ & -53.5$\pm$1.8 & pc \\
             $Z$ & 75.9$\pm$2.5 & pc \\
             $U$ & 6.1$\pm$0.3 & km s$^{-1}$ & \\
             $V$ & -0.57$\pm$0.43 & km s$^{-1}$ & \\
             $W$ & -10.4$\pm$0.5 & km s$^{-1}$ & \\
             \hline
             \end{tabular}}
              \\
              {\textbf{Note}: RV is heliocentric.\\
              $^{a}$epoch J2015.5\\
               \textbf{References}}: \\
              (1) \cite{2011AJ....141...97W}; (2) \cite{2016arXiv161205560C};\\
             (3) \cite{2018arXiv180409365G}; (4) \cite{2017yCat.4034....0H}
\end{table}
%
%
%
\subsection{\textit{Gaia} Astrometry and photometry}
2M1027+0629 is one of the $\sim$1.7 billion sources, which has five-parameter astrometric solution measured by the {\it Gaia} mission \citep{2016A&A...595A...1G}. The {\it Gaia} DR2 \citep{2018arXiv180409365G} released on 25 April, 2018 lists its parallax to be 10.1 $\pm$ 0.3 mas. This corresponds to a distance of 99.0$\pm$2.9 pc, and is farther than the previously reported photometric distance of 67.6 pc \citep{2011AJ....141...97W} asssuming it to be a single star. In addition, the {\it Gaia} proper motions are $\mu_{\alpha}$ = -1.2$\pm$0.5 mas yr$^{-1}$ and $\mu_{\delta}$ = -15.1$\pm$0.5 mas yr$^{-1}$. \\ \\
 \subsection{Spectroscopy}
\begin{figure}
      \includegraphics[scale=0.30,angle=0]{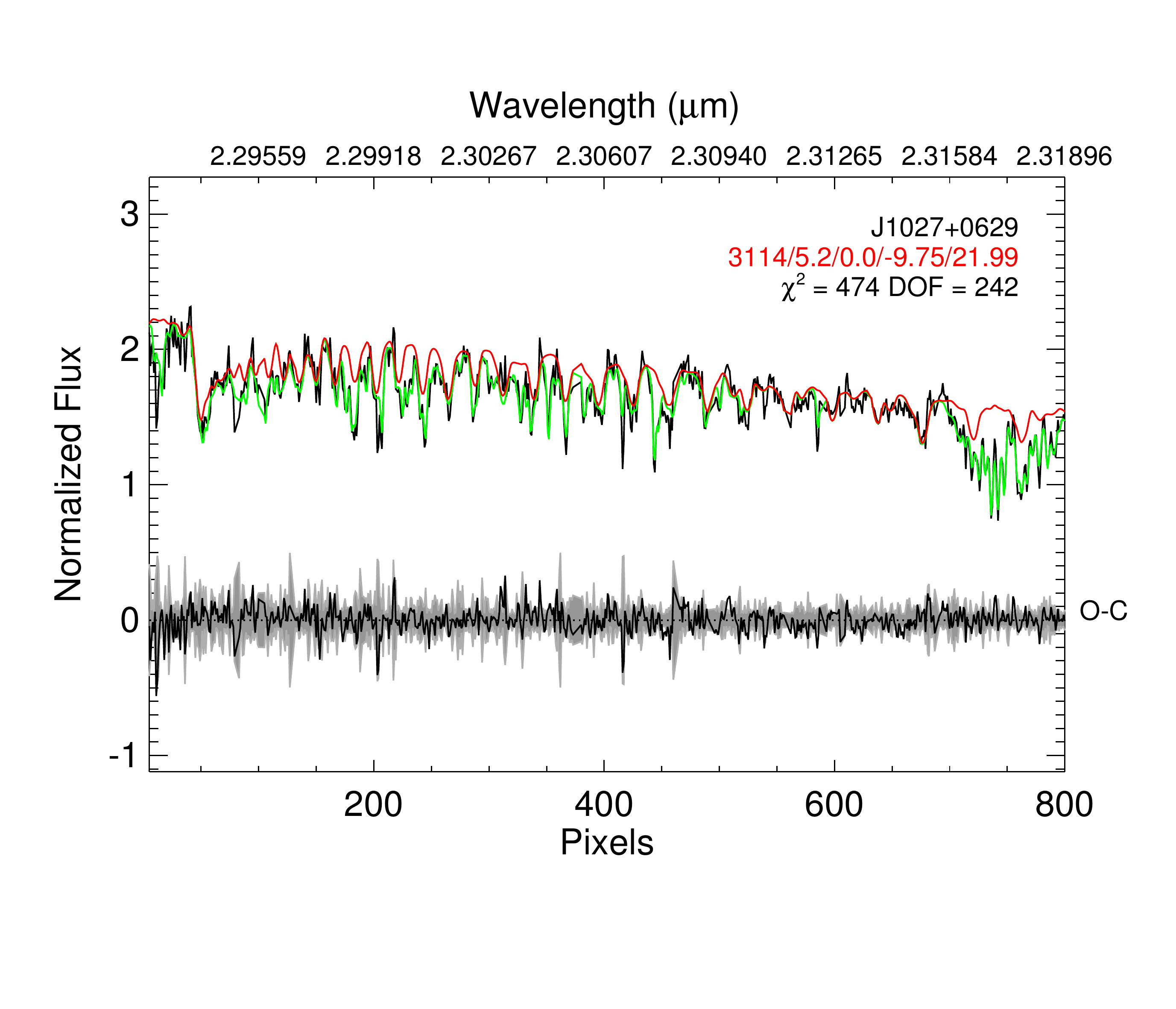} 
       \caption{NIRSPEC spectrum of 2M1027+0629 obtained on UT 2018 April 26 (black line). The red line represents BT-Settl Solar-metallicity spectral model from \citep{2012RSPTA.370.2765A}. The best fit parameters are T$_{eff}$ = 3110$\pm$40 K, log {\it g} = 5.2$\pm$0.2, RV = -9.8$\pm$0.6 km s$^{-1}$ and $v$ sin $i$ = 21.5$\pm$1.1 km s$^{-1}$. The difference between data and model (O-C) is shown in black at the bottom of the plot; the $\pm$1$\sigma$ uncertainty spectrum is indiated in gray.}
       \label{fig:SED}
\end{figure}

 High resolution near-infrared spectra of 2M1027+0629 were obtained with the Near InfraRed SPECtromter (NIRSPEC; \citealt{2000SPIE.4008.1048M}) on the Keck II Telescope on 2018 April 26 (UT) in partly cloudy conditions. The N7 order-sorting filter and 0$\farcs$432-wide slit were used to measure 2.00-2.39 $\mu$m spectra over orders 32-38 with $\lambda/\Delta\lambda$ $\approx$ 20,000 ($\Delta$v$\approx$ 15 km s$^{-1}$) and dispersion of 0.315 {\AA} pixel$^{-1}$. Two exposures of 1500 s each were obtained at an airmass of 1.04, nodding 7$"$ along the slit for sky subtraction. This was followed by observations of the A0 V star 69 Leo (V = 5.404) at similar airmass. Flat field and dark frames were obtained at the start of each night for detector calibration. Data were reduced and the raw extracted spectrum forward modeled following the methodology described in \cite{2016ApJ...827...25B}, with BT-Settl Solar-metallicity spectral models drawn from \cite{2012RSPTA.370.2765A}. The best fit-model (shown in Figure \ref{fig:SED}) has T$_{eff}$ = 3110$\pm$40 K, log {\it g} = 5.2$\pm$0.2 (cgs), RV = -9.8$\pm$0.6 km s$^{-1}$ and $v$ sin $i$ = 21.5$\pm$1.1 km s$^{-1}$. Using radial velocity and {\it Gaia} astrometry ($\alpha$ = 156.940573$^{\circ}$ and $\delta$ = 6.486120$^{\circ}$), $UVW$ components of space velocities of 2M1027+0629 are $U$ = 6.1$\pm$0.3 km s$^{-1}$, $V$ = -0.57$\pm$0.43 km s$^{-1}$, and $W$ = -10.4$\pm$0.5 km s$^{-1}$. \\
%
In Table \ref{table:properties}, we list all the photometric and physical properties of 2M1027+0629, which are available in literature, measured by {\it Gaia} mission, and spectral properties from best-fit model to NIRSPEC spectrum.
\subsection{Flares}
\begin{figure*} 
    \includegraphics[scale=0.60,angle=0]{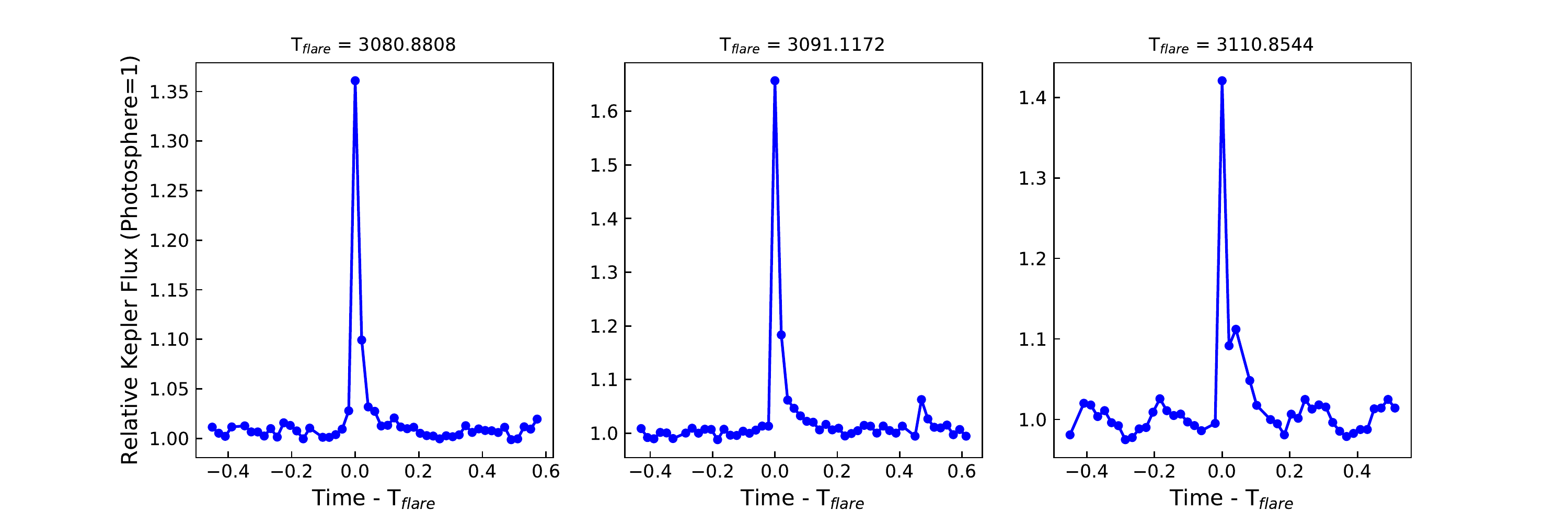} 
    \caption{The three strong flares observed on 2M1027+0629. The blue dots represent the observed data. The time in $X$-axis of each panel is centered at peak flare time mentioned above each panel.}
    \label{fig:flares}
\end{figure*}

We detected three strong white light flares in the $K2$ light curve, as shown in Figure \ref{fig:flares}. The three flares (left to right) in Figure \ref{fig:flares} have equivalent durations (time during which the flare emits the same amount of energy as the star does in its quiescent state) of 15.0 minutes, 29.1 minutes, and 20.3 minutes respectively. To estimate the flare energy, we first calculated the bolometric energy of flare for an equivalent duration of 1 second by approximating the flare to be a 10,000 K hot blackbody and has same count rate through $Kepler$ filter as 2M1027+0629. Assuming that both stars have equal luminosity, same spectral type of M6, and are at same distance of 99.0$\pm$2.9 pc measured by {\it Gaia}, we first estimated bolometric (UV/optical/IR) energy of 10,000 K flare having equivalent duration of 1 second to be 2.9 $\times$ 10$^{30}$ erg. We then multiplied this energy by the equivalent duration to estimate the bolometric (UV/optical/IR) energies of the flares to be 2.6 $\times$ 10$^{33}$, 5.0 $\times$ 10$^{33}$ and 3.5 $\times$ 10$^{33}$ erg respectively. More details about flare energy estimation can be found in \cite{2018ApJ...858...55P}. 
\section{Discussion and Conclusion} \label{sec: discussion}
Using Lomb-Scargle periodogram, we find two closely separated periods 0.2114$\pm$0.0002 days and 0.2199$\pm$0.0003 days, in the $K2$ light curve of 2M1027+0629, and correspond to the rotation period of two stars in a binary system. They form beat patterns in the light curve. The {\it Gaia} parallax  supports the evidence for 2M1207+0629 as a near-equal luminosity binary system.  Previously, \cite{2011AJ....141...97W} reported its photometric distance to be 67.6 pc,  assuming it as a single star. The {\it Gaia} DR2 parallax of  (10.1 $\pm$ 0.3) mas corresponds to (99.0$\pm$2.9) pc, and hence this binary system appears to be twice as luminous as a single M6 star. The values of $UVW$ components of space velocities are consistent with a young age, as are the rotation periods. Furthermore, the membership probabilties calculated by using the BANYAN $\Sigma$ tool \citep{2018ApJ...856...23G} suggest that it does not belong to any known moving group. We also detected three strong white light flares in the $K2$ light curve, which have equivalent durations of 15.0 minutes, 29.1 minutes and 20.3 minutes respectively. We estimated bolometric (UV/optical/IR) energies of those flares to be 2.8 $\times$ 10$^{33}$ erg, 5.2 $\times$ 10$^{33}$ erg, and 3.8 $\times$ 10$^{33}$ erg respectively. If we consider the flares to occur on one of the stars, these flares have amplitudes $\sim$2 relative to the quiescent photospheric level. \\ \\
The beat patterns in the light curves of stars can also arise due to two more phenomena. One is the result of differential rotation of the star, such that spots in different latitudes would have close periods. The horizontal shear differential rotation parameter ($\Delta\Omega$) of M dwarfs are however very small ($<$0.1 rad day$^{-1}$; \citealt{2013A&A...560A...4R}). For the two observed periods here, $\Delta\Omega$ = 1.2 rad day$^{-1}$. The next is the result of stellar pulsations. There is not any observational evidence of pulsating M dwarfs and the theoritical calculations predict pulsations due to convectionally excited oscillations or due to $\epsilon$ mechanism in M dwarfs, have very short periods ranging from a few minutes to about half an hour. In addition, the convectionally excited oscillations in M dwarfs would have amplitudes no more than a few ppm \citep{2012MNRAS.419L..44R,2013MNRAS.430.2313C,2014ApJ...788..114R}. We rule out both possibilities in the case of 2M1207+0629. \\ \\
The two similar rapid rotation periods indicate that both stars are likely to have similar spectral types of M6 \citep{2016AJ....152..115S}. Additional follow-up observation of this system using high resolution AO imaging and spectroscopy is necessary to constrain the separation and the masses of the two stars, and orbital period of the system, which will be valuable to understand the various aspects like rotation-activity-age relationships, evolution of angular momentum in rapidly rotating stars in binary sysem, and gyrochronology. 

\section*{Acknowledgements}
We are thankful to James MacDonald for his helpful suggestions. The material is based upon work supported by NASA under award Nos. NNX15AV64G, NNX16AE55G and NNX16AJ22G. Some/all of the data presented in this paper were obtained from the Mikulski Archive for Space Telescopes (MAST). STScI is operated by the Association of Universities for Research in Astronomy, Inc., under NASA contract NAS5-26555. This paper includes data collected by the Kepler mission. Funding for the Kepler mission is provided by the NASA Science Mission directorate. This work has made use of data from the European Space Agency (ESA) mission
{\it Gaia} (\url{https://www.cosmos.esa.int/gaia}), processed by the {\it Gaia}
Data Processing and Analysis Consortium (DPAC,
\url{https://www.cosmos.esa.int/web/gaia/dpac/consortium}). Funding for the DPAC
has been provided by national institutions, in particular the institutions
participating in the {\it Gaia} Multilateral Agreement. This work made use of the \url{http://gaia-kepler.fun}  crossmatch database created by Megan Bedell.





\bibliographystyle{mnras}
\bibliography{/Users/rishipaudel/GoogleDrive/Research/astrobib}


\bsp	
\label{lastpage}
\end{document}